\documentclass[11 pt, a4paper]{article}
\usepackage{graphicx} 

\usepackage{amsmath}
\usepackage{amssymb}
\usepackage{amsfonts}

\newtheorem{theorem}{Theorem}%
\newtheorem{example}{Example}%
\newtheorem{remark}{Remark}%
\newtheorem{corollary}{Corollary}%
\newtheorem{lemma}{Lemma}%

\newenvironment{proof}{\paragraph{Proof:}}{\hfill$\square$}
\begin{document}
\noindent{\Large{\bf{{An Algorithm to find the Generators of Multidimensional Cyclic Codes over a Finite Chain Ring }}}}\\

\noindent{\textbf{{Disha$\footnote[1]{email: disha.phd21appsc@pec.edu.in} $}}, \textbf{{Sucheta Dutt$\footnote[2]{~*correspondence: sucheta@pec.edu.in }^{,*} $ }}}\\\\
$^{1,2}$Department of Mathematics, Punjab Engineering College, Chandigarh, 160012, India\\\\

\noindent \textbf{Abstract}
The aim of this paper is to determine the
algebraic structure of multidimensional cyclic codes over a finite chain ring $\mathfrak{R}$. An algorithm to find the generator polynomials of $n$ dimensional ($n$D) cyclic codes of length $m_{1}m_{2}\dots m_{n}$ over $\mathfrak{R}$ has been
developed using the generator polynomials of cyclic codes over $\mathfrak{R}$.
Additionally, the generators of $n$D cyclic codes with length $m_{1}m_{2}\dots m_{n}$  over $\mathfrak{R}$ have been obtained as separable polynomials for the case $q\equiv 1(mod~ m_{j}), j\geq 2$, where $q=p^{r}$ is the cardinality of residue field of $\mathfrak{R}$.\\

\noindent \textbf{keywords}
Cyclic codes, $n$D codes, Finite chain ring.

\section{Introduction}\label{sec1}

Contactless transactions driven by customer security, convenience, and efficiency are swiftly opening new opportunities for consumers. Noise in communication systems is an inevitable reality. 
Error-correcting codes have been developed as a solution to this problem. In view of rich algebraic structure and good error correction capability, cyclic codes are among the most studied algebraic codes. $n$ dimensional ($n$D) cyclic codes are an important generalization of cyclic codes.

The basic theory of 2D cyclic codes has been introduced by H. Imai \cite{2} in 1977. Later, the relation between 2D cyclic codes and quasi-cyclic codes has been established by C. Güneri and F. Özbudak \cite{3}. In 2016, 2D cyclic codes of length $n = s2^{k}$ over the finite field $\mathbb{F}_{p^m}$  have been characterized as ideals of the quotient ring $\mathbb{F}_{p^m}[x, y]/\langle x^{s} - 1,y^{2^{k}}- 1 \rangle$ for an odd prime $p$ by Z. Sepasdar and K. Khashyarmanesh \cite{4}. Using a similar approach, the algebraic structure of repeated root 2D constacyclic codes of length $2p^{s}2^{k}$ over a finite field $\mathbb{F}_{p^m}$ has been characterized by Z. Rajabai and K. Khashyarmanesh \cite{5}. The generator matrix of 2D cyclic codes of arbitrary length has been determined by Z. Sepasdar \cite{1} using another approach. The results of \cite{1} have been generalized to 3D and $n$D cyclic codes by R. M. Lalasoa et al. \cite{7}. Recently, using the concept of central primitive idempotents, a new form of generator polynomials of two-dimensional $(\alpha, \beta)$-constacyclic codes of arbitrary length $sl$ has been established by S. Bhardwaj and M. Raka \cite{8}. The results of \cite{8} have been extended to multidimensional constacyclic codes by S. Bhardwaj and M. Raka \cite{9}.

In this paper, the generator polynomials of 2D cyclic codes of arbitrary length over a finite chain ring $\mathfrak{R}$ have been determined by two different methods. Both methods use the structure of cyclic codes over a finite chain ring $\mathfrak{R}$. Method 1 gives the generators of a 2D cyclic code of arbitrary length over $\mathfrak{R}$ as polynomials in $x$ and $y$. Method 2 gives the generators of a 2D cyclic code of length $mn$ such that either $m\vert q-1$ or $n\vert q-1$ as separable polynomials in $x$ and $y$. Moreover, for a natural number $n$, an algorithm based on both the methods for finding generators of an $n$D cyclic code has been developed.

\section{Preliminaries}\label{sec2}

Let $R$ be a finite commutative ring. A code $C$ of length $t$ over $R$ is called a linear code if it is a submodule of $R^t$ over $R$. A linear code $C$ with length $t$ over $R$ is known to be cyclic if $\tau(b) \in C$ for every $b	\in C$, where $\tau$ is the usual cyclic shift operator over $R^t$ defined by $\tau(r_{0},r_{1},\dots,r_{t-1})=(r_{t-1},r_{0},r_{1},\dots,r_{t-2})$. It is well established that a cyclic code $C$ of length $t$ over $R$ can be viewed as an ideal of $R[x]/\langle x^{t}-1\rangle.$  A linear code $\mathcal{C}$ of length $mn$ over $R$  is called a 2D cyclic code  if its codewords, viewed as $m\times n$ arrays of the form
\begin{equation*}
 c = \begin{bmatrix}
    r_{ij}
 \end{bmatrix},~0\leq i\leq m-1,~0\leq j\leq n-1,~r_{ij}\in R 
\end{equation*}
are closed under both row and column cyclic shifts. 

It is easy to check that a 2D cyclic code of length $mn$ over $R$ can be viewed as an ideal of the ring $R[x,y]/\langle x^{m} - 1,y^{n}-1\rangle$. For a natural number $n$, an $n$D cyclic code of length $m_{1}m_{2}\dots m_{n}$ over $R$ is defined as an ideal of the ring $R[x_{1},x_{2},\dots,x_{n} ]/\langle x_{1}^{m_{1}} - 1, x_{2}^{m_{2}} - 1,\dots, x_{n}^{m_{n}} - 1\rangle$. A polynomial $f(x_{1},x_{2},\dots,x_{n} )\in R[x_{1},x_{2},\dots,x_{n} ]$ is said to be separable if $f(x_{1},x_{2},\dots,x_{n} )=f_{1}(x_{1})f_{2}(x_{2})\dots f_{n}(x_{n}) $, where $f_{1}(x_{1}),f_{2}(x_{2}),\dots, f_{n}(x_{n})$ are polynomials over $R$.

If all ideals of a finite commutative ring $R$ form a chain under inclusion operation, then $R$ is said to be a finite chain ring. All ideals of a finite chain ring are principally generated. Moreover, there exists a unique maximal ideal in a finite chain ring. Let $\mathfrak{R}$ be a finite chain ring and $\langle \gamma \rangle$ be its maximal ideal. Let $\nu$ be nilpotency index of $\gamma$ and $\mathbb{F}_{q}=\mathfrak{R}/\langle \gamma \rangle$, where $q=p^{r}.$ Define a map $\bar{}: \mathfrak{R}\longrightarrow \mathbb{F}_{q}$ as $\Bar{r}=r(mod~\gamma)$. Clearly, $\bar{}$ is an onto ring homomorphism. This map can be naturally extended as a map from the polynomial ring $\mathfrak{R}[x]$ to $\mathbb{F}_{q}[x]$.\\\\
The generators of a cyclic code $C$ over a finite chain ring $\mathfrak{R}$ have been determined by Monika et al. \cite{13}. We reproduce below the relevant results from \cite{13}, which we shall require for determining the generator polynomials of a 2D cyclic code over $\mathfrak{R}$. 

\begin{theorem}\label{th.1}\cite{13}
Let $\mathcal{C}$ be a cyclic code over a finite chain ring $\mathfrak{R}$. Then, there exists a positive integer $r$ such that
$\mathcal{C}=\langle p_{0}(x),~p_{1}(x),\dots,~p_{r}(x)\rangle$, $p_j(x)=\gamma^{i_{j}}q_{j}(x)$ and $q_{j}(x)$ is a monic polynomial in $\mathfrak{R}^{j}[x]/\langle x^n-1\rangle$, where $\mathfrak{R}^j$ is the finite chain ring $\mathbb{F}_{q}+\gamma \mathbb{F}_{q}+\gamma^{2} \mathbb{F}_{q}+\dots+ \gamma^{\nu-i_{j}-1}\mathbb{F}_{q}$ for $0\leq j\leq r$. Also, $i_{0}>i_{1}>\dots >i_{r}$ and $t_{0}<t_{1}<\dots<t_{r}$, where $t_j=deg(p_{j}(x)).$
\end{theorem}

\begin{corollary}\label{cor.1}\cite{13}
A cyclic code of arbitrary length over $\mathfrak{R}$ is generated by at most $k=min\{\nu, t_{r}+1\}$ elements.
\end{corollary}

\section{Generators of 2D cyclic codes of length $mn$  over a finite chain ring}\label{sec3}
\subsection{Method 1}\label{subsec2}
Let $\mathcal{C}$ be a 2D cyclic code of length $mn$ over a finite chain ring $\mathfrak{R}$. Then, $\mathcal{C}$ can be viewed as an ideal of the ring $\mathfrak{R}[x,~ y]/ \langle x^{m}-1,~ y^{n}-1\rangle$ which can be easily seen to be isomorphic to $(\mathfrak{R}[x]/ \langle x^{m}- 1 \rangle)[y]/\langle y^{n}-1\rangle$ under the map $\displaystyle\sum_{i=0}^{m-1} \displaystyle\sum_{j=0}^{n-1} a_{ij} x^{i} y^j \longrightarrow  \displaystyle\sum_{j=0}^{n-1}\Big{(} \displaystyle\sum_{i=0}^{m-1} a_{ij} x^{i}\Big{)} y^j $. Let $f(x,y)\in \mathcal{C}$ be any element. Then, $f(x,y)$ can be uniquely written as $\displaystyle\sum_{j=0}^{n-1} f_{j}(x)y^{j}, \text{ where } f_{j}(x)\in \mathfrak{R}[x]/ \langle x^{m}- 1 \rangle \text{ for } 0\leq j \leq {n- 1}$. 
Define, $I_{j}= \big{\{} g_{j}(x)\in \mathfrak{R}[x]/ \langle x^{m}- 1 \rangle\mid \exists~ g(x,y)\in \mathcal{C}~\text{such that}~ g(x,y)=\displaystyle\sum_{k=0}^{n-1-j} g_{n-1-k}(x)y^{k}\big{\}}$, $0\leq j\leq n-1$.
It can be easily verified that each $I_{j},~0\leq j\leq n-1$, is an ideal of the ring $\mathfrak{R}[x]/ \langle x^{m}- 1 \rangle$ and therefore a cyclic code of length $m$ over $\mathfrak{R}$. By Theorem \ref{th.1}, we can find a set of polynomials $p_{0}^{(j)}(x),~p_{1}^{(j)}(x)\dots,~p_{r_{j}}^{(j)}(x)\in \mathfrak{R}[x]/ \langle x^{m}- 1 \rangle$ such that 
\begin{align*}
I_{j}= \langle p_{0}^{(j)}(x),~p_{1}^{(j)}(x)\dots,~p_{r_{j}}^{(j)}(x)\rangle,~0\leq j\leq n-1.
\end{align*}
Also, by Corollary \ref{cor.1}, $r_{j}+1\leq min(\nu,~t_{r_{j}}+1)$ for each $j,~0\leq j\leq n-1$, where $t_{r_{j}}= deg(p_{r_{j}}^{(j)}(x))$.

\begin{theorem}\label{th.2}
Let $\mathcal{C}$ be a $2D$ cyclic code of length $mn$ over $\mathfrak{R}$. Then the set $\{P_{i}^{(j)}(x,y)\in \mathfrak{R}[x,y]\mid0\leq i\leq {r_{j}},~0\leq j\leq {n-1}\}$ generates $\mathcal{C}$ where 
\begin{align*}
    P_{i}^{(j)}(x,y)&=\displaystyle\sum_{k=0}^{n-1-j} a_{ik}^{(j)}(x)y^{k},~\text{for all }~0\leq j\leq {n-1} 
    \end{align*}
    and $a_{ik}^{(j)}(x)\in I_{j}$ for every $0\leq k\leq {n-1-j}$. Also, $a_{i(n-1-j)}^{(j)}(x)=p_{i}^{(j)}(x),~0\leq i \leq r_{j}, ~ 0\leq j\leq {n-1}$.
\end{theorem}

\begin{proof}
Let $\mathcal{C}$ be a 2D cyclic code of length $mn$ over $\mathfrak{R}$. Let $ f(x, y) =\displaystyle\sum_{j=0}^{n-1} f_{j}(x)y^{j}$ with $f_{j}(x)\in \mathfrak{R}[x]/ \langle x^{m}- 1 \rangle,~ 0\leq j \leq {n- 1}$, be any element of $\mathcal{C}$. 
Clearly, $f_{n-1}(x)\in I_{0}$. Therefore,
\begin{align*}
 f_{n-1}(x)= \displaystyle\sum_{i=0}^{r_{0}} p_{i}^{(0)}(x)t_{i}^{(0)}(x),~\text{where}~ t_{i}^{(0)}(x)\in \mathfrak{R}[x]/ \langle x^{m}- 1 \rangle~\text{for all}~ 0\leq i\leq {r_{0}} 
 \end{align*}
Also, $p_{i}^0(x)\in I_{0};~0\leq i \leq {r_{0}} $, therefore by definition of $I_{0}$, there exists ${P_{i}^{(0)}(x,y)\in \mathcal{C}}$ such that
\begin{align*}
P_{i}^{(0)}(x,y)=\displaystyle\sum_{k=0}^{n-1} a_{ik}^{(0)}(x)y^{k},~\text{where}~ a_{i(n-1)}^{(0)}(x)=p_{i}^{(0)}(x),~0\leq i \leq {r_{0}}.
\end{align*}
Since $\mathcal{C}$ is an ideal of $\mathfrak{R}$, $y^{j}P_{i}^{(0)}(x,y)\in \mathcal{C}$. Therefore,  $a_{ik}^{(0)}(x)\in I_{0}$ for all $0\leq k\leq {n-1},~0\leq i \leq {r_{0}}.$ Let 
\begin{align}
h_{1}(x, y)&= f(x, y)-\displaystyle\sum_{i=0}^{r_{0}}P_{i}^{(0)}(x,y)t_{i}^{(0)}(x)\\ \nonumber
&=\displaystyle\sum_{k=0}^{n-2} f_{k}(x)y^{k} -\displaystyle\sum_{i=0}^{r_{0}}t_{i}^{(0)}(x)\big{(}\displaystyle\sum_{k=0}^{n-2} a_{ik}^{(0)}(x)y^{k}\big{)}\\ \nonumber
&=\displaystyle\sum_{k=0}^{n-2} h_{k}^{(1)}(x)y^{k};~ h_{k}^{(1)}(x)\in \mathfrak{R}[x]/ \langle x^{m}- 1 \rangle.
\end{align}
Clearly, $h_{1}(x,y)\in \mathcal{C}$ since $f(x,y),P_{i}^{(0)}(x,y)\in \mathcal{C}$ for every $0\leq i\leq{r_{0}}$. Also, by definition of $I_{1}$,  $h_{n-2}^{(1)}(x)\in I_{1}$. So,
\begin{align*}
 h_{n-2}^{(1)}(x)= \displaystyle\sum_{i=0}^{r_{1}}p_{i}^{(1)}(x)t_{i}^{(1)}(x),~\text{where}~ t_{i}^{(1)}(x)\in \mathfrak{R}[x]/ \langle x^{m}- 1 \rangle,~ \text{for all}~ 0\leq i\leq {r_{1}}
 \end{align*}
Now, $p_{i}^{(1)}(x)\in I_{1};~0\leq i \leq {r_{1}} $, therefore there exists ${P_{i}^{(1)}(x,y)\in \mathcal{C}}$ such that
\begin{align*}
P_{i}^{(1)}(x,y)=\displaystyle\sum_{k=0}^{n-2} a_{ik}^{(1)}(x)y^{k},~\text{where}~ a_{i(n-2)}^{(1)}(x)=p_{i}^{(1)}(x),~0\leq i \leq {r_{1}}.
\end{align*}
Also, since $\mathcal{C}$ is an ideal of $\mathfrak{R}$, $y^{j}P_{i}^{(1)}(x,y)\in \mathcal{C}$. Therefore $a_{ik}^{(1)}(x)\in I_{1}$ for all $0\leq k\leq {n-2},~0\leq i \leq {r_{1}}.$ Let
\begin{align}
h_{2}(x,y)&= h_{1}(x,y)-\displaystyle\sum_{i=0}^{r_{1}}P_{i}^{(1)}(x,y)t_{i}^{(1)}(x)\\ \nonumber
&=\displaystyle\sum_{k=0}^{n-3} h_{k}^{(1)}(x)y^{k} -\displaystyle\sum_{i=0}^{r_{1}}t_{i}^{(1)}(x)\big{(}\displaystyle\sum_{k=0}^{n-3} a_{ik}^{(1)}(x)y^{k}\big{)}\\\nonumber
&=\displaystyle\sum_{k=0}^{n-3} h_{k}^{(2)}(x)y^{k};~ h_{k}^{(2)}(x)\in \mathfrak{R}[x]/ \langle x^{m}- 1 \rangle.
\end{align}
Clearly, $h_{2}(x,y)\in \mathcal{C}$ since $h_{1}(x,y),~P_{i}^{(1)}(x,y)\in \mathcal{C};~ 0\leq i\leq{r_{1}}$. Continuing in this fashion, we obtain polynomials $h_{1}(x,y),~h_{2}(x,y),\dots, h_{n-1}(x,y)$, where
\begin{align*}
h_{n-1}(x,y)= h_{n-2}(x, y)-\displaystyle\sum_{i=0}^{r_{n-2}}P_{i}^{(n-2)}(x,y)t_{i}^{(n-2)}(x)=h_{0}^{(n-2)}(x).\tag{n-1}
\end{align*}
Clearly, by definition of $I_{n-1}$, $h_{0}^{(n-2)}(x)\in I_{n-1}$, therefore
\begin{align*}
h_{0}^{(n-2)}(x)=\displaystyle\sum_{i=0}^{r_{n-1}}p_{i}^{(n-1)}(x)t_{i}^{(n-1)}(x),~\text{where}~ t_{i}^{(n-1)}(x)\in \mathfrak{R}[x]/ \langle x^{m}- 1 \rangle;~ 0\leq i \leq {r_{n-1}}.
 \end{align*}
Now $p_{i}^{(n-1)}(x)\in I_{n-1}$, therefore there exist
$ {P_{i}^{(n-1)}(x,y)\in \mathcal{C}}$ such that
\begin{align*}
P_{i}^{(n-1)}(x,y)=p_{i}^{(n-1)}(x)
\end{align*}
Therefore,
\begin{align*}
    h_{n-1}(x)=h_{0}^{(n-2)}(x)= \displaystyle\sum_{i=0}^{r_{n-1}}P_{i}^{(n-1)}(x,y)t_{i}^{(n-1)}(x)\tag{n}
\end{align*}
It follows from equations (1) to (n) that
\begin{align*}
f(x, y)&=\displaystyle\sum_{i=0}^{r_{0}}P_{i}^{(0)}(x,y)t_{i}^{(0)}(x)+
\displaystyle\sum_{i=0}^{r_{1}}P_{i}^{(1)}(x,y)t_{i}^{(1)}(x)\\
&+
\dots+\displaystyle\sum_{i=0}^{r_{n-2}}P_{i}^{(n-2)}(x,y)t_{i}^{(n-2)}(x)+
\displaystyle\sum_{i=0}^{r_{n-1}}P_{i}^{(n-1)}(x,y)t_{i}^{(n-1)}(x)
\end{align*}
Thus, the set $\{P_{i}^{(j)}(x,y)\in \mathfrak{R}[x,y]\mid 0\leq i\leq {r_{j}},~0\leq j\leq {n-1}\}$ generates $\mathcal{C}$,
 where 
\begin{align*}
    P_{i}^{(j)}(x,y)&=\displaystyle\sum_{k=0}^{n-1-j} a_{ik}^{(j)}(x)y^{k},~\text{for all }~0\leq j\leq {n-1} 
    \end{align*}
    and $a_{ik}^{(j)}(x)\in I_{j}$ for every $0\leq k\leq {n-1-j}$. Also, $a_{i(n-1-j)}^{(j)}(x)=p_{i}^{(j)}(x),~0\leq i \leq r_{j}, ~ 0\leq j\leq {n-1}$.
\end{proof}
The following result is an immediate consequence of Theorem \ref{th.2} and Corollary \ref{cor.1}.

\begin{corollary}\label{cor.2}
The number of generators of a 2D cyclic code $\mathcal{C}$ over $\mathfrak{R}$ is at most $kn$ where $k=\displaystyle\sum_{j=0}^{n-1}(r_{j}+1)$.
\end{corollary}

\begin{remark}\label{r1}
A 2D cyclic code of length $mn$ can also be viewed as a 2D cyclic code of length $nm$ by interchanging the rows and columns of its codewords which is clear from the following ring isomorphisms
\begin{align*}
    \mathfrak{R}[x,~y]/\langle x^{m}-1,~y^{n}-1\rangle &\cong (\mathfrak{R}[x]/\langle x^{m}-1\rangle)[y]/\langle y^{n}-1\rangle\\
    &\cong (\mathfrak{R}[y]/\langle y^{n}-1\rangle)[x]/ \langle x^{m}-1\rangle
\end{align*}
\end{remark}

\begin{remark}\label{r2}
In view of Remark \ref{r1}, we may work with ideals of $(\mathfrak{R}[y]/\langle y^{n}-1\rangle)[x]/ \langle x^{m}-1\rangle $ instead of $(\mathfrak{R}[x]/\langle x^{m}-1\rangle)[y]/\langle y^{n}-1\rangle$ to find the generators of a 2D cyclic code $\mathcal{C}$ and may consequently obtain a simpler set of generators of $\mathcal{C}$.
\end{remark}

\begin{example}
Consider the finite chain ring $\mathfrak{R}= \mathbb{F}_{4}+\gamma \mathbb{F}_{4};~ \gamma^{2}=0$. Let $\mathcal{C}$ be 2D cyclic code of length $mn$ over $\mathfrak{R}$, where $m=8~ \text{and} ~n=3$. Then $\mathcal{C}$ can be viewed as an ideal of $\mathfrak{R}[x,y]/\langle x^{8}-1,~y^{3}-1\rangle$. 
Consider the following cyclic codes of length $8$ over $\mathfrak{R}$ given by
\begin{align*}
    I_{0}&=\langle p_{0}^{(0)}(x),~p_{1}^{(0)}(x)\rangle= \langle \gamma(x^3-3x^{2}+3x-1),~x^{4}-4x^{3}+6x^{2}-4x+1\rangle\\
    I_{1}&=\langle p_{0}^{(1)}(x)\rangle=\langle \gamma(x^{4}-4x^{3}+6x^{2}-4x+1)\rangle\\
     I_{2}&=\langle p_{0}^{(2)}(x)\rangle =\langle x^{6}-6x^{5}+15x^{4}-20x^{3}+15x^{2}-6x+1 \rangle
\end{align*}
By Theorem \ref{th.2}, the set $\{P_{0}^{(0)}(x,y),~P_{1}^{(0)}(x,y),~P_{0}^{(1)}(x,y),~P_{0}^{(2)}(x,y)\}$ generates $\mathcal{C}$, where 
\begin{align*}
    P_{0}^{(0)}(x,y)&= p_{00}^{(0)}(x)+p_{01}^{(0)}(x)y+\gamma(x^3-3x^{2}+3x-1)y^{2}\\
    P_{1}^{(0)}(x,y)&=p_{10}^{(0)}(x)+p_{11}^{(0)}(x)y+(x^{4}-4x^{3}+6x^{2}-4x+1)y^{2}\\
    P_{0}^{(1)}(x,y)&=p_{00}^{(1)}(x)+\gamma(x^{4}-4x^{3}+6x^{2}-4x+1)y\\
    P_{0}^{(2)}(x,y)&=(x^{6}-6x^{5}+15x^{4}-20x^{3}+15x^{2}-6x+1)
\end{align*}
such that $p_{00}^{(0)}(x),~p_{01}^{(0)}(x),~p_{10}^{(0)}(x),~p_{11}^{(0)}(x)\in I_{0},~p_{00}^{(1)}(x)\in I_{1}$. More specifically, taking $p_{00}^{(0)}(x)=p_{01}^{(0)}(x)=p_{10}^{(0)}(x)=x^{4}-4x^{3}+6x^{2}-4x+1~ \text{and}~ p_{11}^{(0)}(x)= \gamma(x^3-3x^{2}+3x-1),~p_{00}^{(1)}(x)= \gamma(x^{4}-4x^{3}+6x^{2}-4x+1)$ we get, the set $\big{\{} (x^{4}-4x^{3}+6x^{2}-4x+1)(1+y)+\gamma(x^3-3x^{2}+3x-1)y^{2}, (x^{4}-4x^{3}+6x^{2}-4x+1)+\gamma(x^3-3x^{2}+3x-1)y+(x^{4}-4x^{3}+6x^{2}-4x+1)y^{2}, \gamma(x^{4}-4x^{3}+6x^{2}-4x+1)+\gamma(x^{4}-4x^{3}+6x^{2}-4x+1)y, (x^{6}-6x^{5}+15x^{4}-20x^{3}+15x^{2}-6x+1)\big{\}}$, which generates a 2D cyclic code of length $mn$ over $\mathfrak{R}$.
\end{example}
\subsection{Method 2}\label{subsec4}

Generator polynomials of 2D $(\alpha, \beta)$-constacyclic codes of arbitrary length $s.l$ with $q\equiv 1( mod ~l)$ over finite fields $\mathbb{F}_{q}$ have been determined in separable form by S. Bhardwaj and M. Raka \cite{8}.  In this subsection, we use their approach to get generators of 2D cyclic codes of length $mn$ where, $q\equiv 1( mod ~n)$ over a finite chain ring $\mathfrak{R}$ with residue field $\mathbb{F}_{q}$, in separable form.\\\\
We shall require following results for later use.
\begin{theorem}\label{th.3}\cite{14}
Let $g(x)\in \mathfrak{R}[X]$ be a monic polynomial such that $\Bar{g}(x)= f_{1}(x)f_{2}(x)\dots f_{m}(x)$, where $f_{i}(x) \in \mathbb{F}_q[X]$ are pairwise coprime monic  polynomials for $1\leq i\leq m$. Then there exist monic, pairwise coprime polynomials $g_{i}(x) \in \mathfrak{R}[X]$ such that $g(x)= g_{1}(x)g_{2}(x)\dots g_{m}(x)$.

\end{theorem}
\begin{theorem}\label{th.4}\cite{14}
Let $\mathfrak{R}$ be a finite chain ring and $g(x)\in \mathfrak{R}[x]$ be a monic polynomial. Then $g(x)$ factors uniquely in $\mathfrak{R}[x]$ if $\Bar{g}(x)$ is square free.
\end{theorem}
\begin{lemma}\label{lem.1}
Let $\mathfrak{R}$ be a finite chain ring. Then the following hold in $\mathfrak{R}[y]/\langle y^{n}-1\rangle$.
\begin{enumerate}
\label{lem. 1(a)}\item[(a)] There exists an element $\zeta \in \mathfrak{R}$ which is a primitive $n^{th}$ root of unity.
\label{l1.1}\item[(b)] $\theta_{i}(y)= \frac{1}{n}(1+\zeta^{n-i}y+(\zeta^{n-i}y)^{2}+\dots+ (\zeta^{n-i}y)^{n-1})~\text{for} ~0\leq i\leq n-1$
are primitive central idempotents of $\mathfrak{R}[y]/\langle y^{n}-1\rangle$.
\label{l1.2}\item[(c)]$\theta_{i}(y)y^{j}=(\zeta^{i})^{j}\theta_{i}(y)$ for $0\leq i,~j\leq n-1$.\\
\end{enumerate} 
\end{lemma}

\begin{proof}
\begin{enumerate}
\item[(a)]
Consider the ring $\mathbb{F}_{q}[y]/\langle y^{n}-1\rangle$; where $\mathbb{F}_{q}$ is the residue field of $\mathfrak{R}$ and $q\equiv 1 (modn)$. Let $\omega\in \mathbb{F}_{q}$ be primitive $n^{th}$ root of unity. Then
\begin{align*}
    y^{n}-1= (y-1)(y-\omega)(y-\omega^{2})\dots (y-\omega^{n-1})~ \text{in} ~\mathbb{F}_{q}(y)
\end{align*}
Let $\Bar{\zeta}= \omega$ for some $\zeta\in \mathfrak{R}$. By Theorem \ref{th.3} and Theorem 4, it is easy to see that $y^{n}-1= (y-1)(y-\zeta)(y-\zeta^{2})\dots (y-\zeta^{n-1}) ~ \text{in} ~\mathfrak{R}_{q}(y)$, so that $\zeta$ is a primitive $n^{th}$ root of unity in $\mathfrak{R}$.
\item[(b)]
    $\theta_{i}(y)=\frac{1}{n}(1+\zeta^{n-i}y+(\zeta^{n-i}y)^{2}+\dots+ (\zeta^{n-i}y)^{n-1})\\
~~~~~~~=\frac{1}{n\zeta^{-i}}\big{(}\zeta^{-i}(1+\zeta^{n-i}y+(\zeta^{n-i}y)^{2}+\dots+ (\zeta^{n-i}y)^{n-1})\big{)}\\
~~~~~~~=\prod_{j\neq i} \frac{(y-\zeta^{j})}{(\zeta^i-\zeta^j)}~\text{for}~ 0\leq i\leq {n-1}.$\\
It follows that for $i\neq j$, $\theta_{i}(y)\theta_{j}(y)= (y^{n}-1)g(y)~\text{for some}~ g(y)\in \mathfrak{R}[y]/\langle y^{n}-1\rangle$ and therefore $\theta_{i}(y)\theta_{j}(y)=0$.
Further, it is easy to see that $y^{n}-1$ is a factor of $\theta_{i}(y)(\theta_{i}(y)-1)$. Hence, $\theta_{i}(y)(\theta_{i}(y)-1)=0$ in $\mathfrak{R}[y]/\langle y^{n}-1\rangle$ and therefore, we have $\theta_{i}^{2}(y)=\theta_{i}(y)$. Also, $\theta_{0}(\zeta^{i})+\theta_{1}(\zeta^{i})+\dots+\theta_{n-1}(\zeta^{i})=1,~0\leq i\leq n-1$, implies that $(y-\zeta^{i})\mid \theta_{0}(y)+\theta_{1}(y)+\dots+\theta_{n-1}(y)-1$ which further implies that $\theta_{0}(y)+\theta_{1}(y)+\dots+\theta_{n-1}(y)=1$ in $\mathfrak{R}[y]/\langle y^{n}-1\rangle$.
Hence, $\theta_{i}(y),~ 0\leq i\leq {n-1}$ are primitive central idempotents in $\mathfrak{R}[y]/\langle y^{n}-1\rangle$.
\item[(c)] 
    $\theta_{i}(y)y=\frac{1}{n}(y+\zeta^{n-i}y^{2}+\zeta^{2{n-i}}y^{3}+\dots+\zeta^{(n-i)(n-1)}y^{n})\\
    ~~~~~~~~~=\frac{1}{n}(\zeta^{i}+y+\zeta^{n-i}y^{2}+\dots+ \zeta^{(n-i)(n-2)}y^{n-1})\\
    ~~~~~~~~~=\frac{\zeta^{i}}{n}(1+\zeta^{n-i}y+(\zeta^{n-i}y)^{2}+\dots+ (\zeta^{n-i}y)^{n-1})= \zeta^{i}\theta_{i}(y)$\\
    
Hence $\theta_{i}(y)y^{j}=(\zeta^{i})^{j}\theta_{i}(y)$ for $0\leq i,~j\leq n-1$.
\end{enumerate}
\end{proof}
\begin{remark}
In the proof of the Lemma \ref{lem. 1(a)} above, we have used Theorem \ref{th.3} and Theorem \ref{th.4} to find a primitive $n^{th}$ root of unity $\zeta$ in $\mathfrak{R}$. However, it can be easily checked that if $\omega$ is primitive $n^{th}$ root of unity in the residue field $\mathbb{F}_{q}$ of $\mathfrak{R}$, then $\zeta=\omega^{\gamma^{\nu-1}}$ is primitive $n^{th}$ root of unity over $\mathfrak{R}$. 
\end{remark}
Now, define the sets
    $C_{j}= \{g_{j}(x)\in \mathfrak{R}[x]/\langle x^{m}-1\rangle\mid g_{j}(x)\theta_{j}(y)\in \mathcal{C}\};~
0\leq j\leq {n-1}$. 
It can be easily verified that each $C_{j},~0\leq j\leq n-1$, is an ideal of the ring $\mathfrak{R}[x]/ \langle x^{m}- 1 \rangle$ and therefore a cyclic code of length $m$ over $\mathfrak{R}$. By Theorem \ref{th.1}, we can find a set of polynomials $p_{0}^{(j)}(x),~p_{1}^{(j)}(x)\dots,~p_{r_{j}}^{(j)}(x)\in \mathfrak{R}[x]/ \langle x^{m}- 1 \rangle$ such that 
\begin{align*}
C_{j}= \langle p_{0}^{(j)}(x),~p_{1}^{(j)}(x)\dots,~p_{r_{j}}^{(j)}(x)\rangle;~ 0\leq j\leq n-1
\end{align*}
Also, by Corollary \ref{cor.1}, $r_{j}+1\leq min(\nu,~t_{r_{j}})$ for each $j$, $0\leq j\leq n-1$, where $t_{r_{j}}= deg(p_{r_{j}}^{(j)}(x))$.
\begin{theorem}\label{th.5}
Let $\mathcal{C}$ be a cyclic code of length $mn$ over a finite chain ring $\mathfrak{R}$  with residue field $\mathbb{F}_{q}$ such that $q\equiv 1 (mod~n)$. Then the set $\{\theta_{j}(y)p_{i}^{(j)}(x)\mid 0\leq i\leq {r_{j}},~0\leq j\leq {n-1}\}$ generates $\mathcal{C}$
where, $\theta_{j}(y)$ are primitive central idempotents of $\mathfrak{R}[y]/\langle y^{n}-1\rangle$ as obtained in Lemma \ref{lem.1}.
\end{theorem}

\begin{proof}
Let $\mathcal{C}$ be a 2D cyclic code over $\mathfrak{R}$ and $f(x,y)\in \mathcal{C}$ be any element. Then, 
\begin{align*}
\theta_{j}(y)f(x,y)&= \theta_{j}(y)[f_{0}(x)+f_1(x)y+\dots+f_{n-1}(x)y^{n-1}]\\
&=\theta_{j}(y)f_{0}(x)+\theta_{j}(y)f_1(x)y+\dots+\theta_{j}(y)f_{n-1}(x)y^{n-1}\\
&=\theta_{j}(y)f_{0}(x)+\zeta^{j}\theta_{j}(y)f_1(x)+\dots+(\zeta^{j})^{n-1}\theta_{j}(y)f_{n-1}(x) ~\text{By Lemma \ref{l1.2}}\\
&=\theta_{j}(y)f(x,\zeta^{j})
\end{align*}
By definition of $C_{j}$, $f(x,\zeta^{j})\in C_{j},~0\leq j\leq  n-1$. Therefore, $f(x,\zeta^{j})=p_{0}^{(j)}(x)t_{0}^{(j)}(x)+p_{1}^{(j)}(x)t_{1}^{(j)}(x)+\dots+p_{r_{j}}^{(j)}(x)t_{r_{j}}^{(j)}(x)$ for some  $t_{i}^{(j)}(x)\in \mathfrak{R}[x]/\langle x^{m}-1\rangle$.
Also, we have
\begin{align*}
    f(x,y)&=f(x,y)\displaystyle\sum_{j=0}^{n-1} \theta_{j}(y)=\displaystyle\sum_{j=0}^{n-1} \theta_{j}(y)f(x,~\zeta^{j})\\
    &=\displaystyle\sum_{j=0}^{n-1} \theta_{i}(y)[p_{0}^{(j)}(x)t_{0}^{(j)}(x)+p_{1}^{(j)}(x)t_{1}^{(j)}(x)+\dots+p_{r_{j}}^{(j)}(x)t_{r_{j}}^{(j)}(x)]
\end{align*}
Thus, the set $\{\theta_{j}(y)p_{i}^{(j)}(x)\mid 0\leq i\leq {r_{j}},~0\leq j\leq {n-1}\}$ generates $\mathcal{C}$.
\end{proof}

\begin{example}
Consider the finite chain ring $\mathfrak{R}= \mathbb{Z}_{25}$ with residue field $\mathbb{F}_{5}$ and nilpotency index 2. Let $\mathcal{C}$ be 2D cyclic code of length $mn$ over $\mathfrak{R}$, where $m=10~ \text{and} ~n=4$. Then $\mathcal{C}$ can be viewed as an ideal of $R=\mathfrak{R}[x,y]/\langle x^{10}-1,~y^4-1\rangle$. It can be easily seen that $7$ is primitive $4^{th}$ root of unity in $\mathbb{Z}_{25}$. Therefore, by Lemma \ref{lem.1}, $\theta_{0}(y)= 19(1+y+y^{2}+y^{3}), \theta_{1}(y)= 19(1+18y-y^{2}+7y^{3}), \theta_{2}(y)= 19(1-y+y^{2}-y^{3}), \theta_{3}(y)= 19(1+7y-y^{2}+18y^{3})$
are primitive central idempotents of $\mathfrak{R}[y]/\langle y^{4}-1\rangle$. Consider the following cyclic codes of length $10$ over $\mathfrak{R}$
\begin{align*}
    I_{0}&= I_{2} = \langle p_{0}^{(0)}(x),~p_{1}^{(0)}(x)\rangle\\ I_{1}&= I_{3}=\langle p_{0}^{(1)}(x)\rangle 
\end{align*}
 where $p_{0}^{(0)}(x)=5q_{0}^{(0)}(x)= 5(x^8+x^6+x^4+x^2+1),~p_{1}^{(0)}(x)= q_{1}^{(0)}(x)= x^9+x^8+x^7+x^6+x^5+x^4+x^3+x^2+x+1 ~\text{and}~ p_{0}^{(1)}(x)=x-1$. By Theorem \ref{th.5}, the set
$\{ \theta_{0}(y)p_{0}^{(0)}(x),~\theta_{0}(y)p_{1}^{(0)}(x),~\theta_{1}(y)p_{0}^{(1)}(x),~\theta_{2}(y)p_{0}^{(0)}(x),~\theta_{2}(y)p_{1}^{(0)}(x),~\theta_{3}(y)p_{0}^{(1)}(x)\}$ generates a 2D cyclic code over $\mathfrak{R}$.
\end{example}

\begin{example}
Consider the finite chain ring $\mathfrak{R}= \mathbb{F}_{13}+\gamma \mathbb{F}_{13};~ \gamma^{2}=0$. Let $\mathcal{C}$ be 2D cyclic code of length $mn$ over $\mathfrak{R}$, where $m=169~ \text{and} ~n=12$. Then $\mathcal{C}$ can be viewed as an ideal of $R=\mathfrak{R}[x,y]/\langle x^{169}-1,~y^{12}-1\rangle$. It can be easily seen that $2^{\gamma}(mod~169)$ is primitive $12^{th}$ root of unity in $\mathfrak{R}$. Therefore, by Lemma \ref{lem.1}, $\theta_{j}(y)= \frac{1}{12}(1+2^{\gamma(12-j)} y+2^{2\gamma(12-j)}y^{2}+\dots + 2^{11{\gamma(12-j)}}y^{11})$
are primitive central idempotents of $\mathfrak{R}[y]/\langle y^{12}-1\rangle;~ 0\leq j\leq 11$. Consider the following cyclic codes of length $169$ over $\mathfrak{R}$
\begin{align*}
    I_{0}&=I_{1}=\langle p_{0}^{(0)}(x)\rangle= \langle x^{5}-5x^{4}+10x^{3}-10x^{2}+5x-1\rangle\\ 
    I_{2}&=\langle p_{0}^{(2)}(x)\rangle=\langle x^{6}-6x^{5}+15x^{4}-20x^{3}+15x^{2}-6x+1\rangle\\
    I_{3}&=\langle p_{0}^{(3)}(x)\rangle=\langle \gamma(x^{7}-7x^{6}+21x^{5}-35x^{4}+35x^{3}-21x^{2}+7x-1)\rangle\\
    I_{4}&=\langle p_{0}^{(4)}(x)\rangle=\langle \gamma(x^{8}-8x^{7}+28x^{6}-56x^{5}+70x^{4}-56x^{3}+28x^{2}-8x+1)\rangle\\
     I_{5}&=I_{6}=\langle p_{0}^{(5)}(x),~p_{1}^{(5)}(x)\rangle\\
     &=\langle\gamma(x^{4}-4x^{3}+6x^{2}-4x+1),\\
     &~~~~~x^{5}-5x^{4}+10x^{3}-10x^{2}+5x-1+\gamma(x^{3}-3x^{2}+4x-2) \rangle\\
    I_{7}&=I_{8}=\langle p_{0}^{(7)}(x)\rangle\\
    &=\langle \gamma(x^{9}-9x^{8}+36x^{7}-84x^{6}+126x^{5}-126x^{4}+84x^{3}-36x^{2}+9x-1)\rangle\\
    I_{9}&=I_{10}=\langle p_{0}^{(9)}(x)\rangle=\langle \gamma(x^{6}-6x^{5}+15x^{4}-20x^{3}+15x^{2}-6x+1) \rangle\\
    I_{11}&=\langle p_{0}^{(11)}(x),~p_{1}^{(11)}(x)\rangle=\langle \gamma(x^{6}+126x^{5}-126x^{4}+84x^{3}-36x^{2}+9x-1),\\
    &~~~~~~~~~~~~~~~~~~ x^{8}-8x^{7}+28x^{6}-56x^{5}+70x^{4}-56x^{3}+28x^{2}-8x+1 \rangle
\end{align*}
 By Theorem \ref{th.5}, the set $\{ \theta_{0}(y)p_{0}^{(0)}(x),~\theta_{1}(y)p_{0}^{(0)}(x),~\theta_{2}(y)p_{0}^{(2)}(x),~\theta_{3}(y)p_{0}^{(3)}(x),\\
 \theta_{4}(y)p_{0}^{(4)}(x),~
 \theta_{5}(y)p_{0}^{(5)}(x), \theta_{5}(y)p_{1}^{(5)}(x),~\theta_{6}(y)p_{0}^{(5)}(x),~\theta_{6}(y)p_{1}^{(5)}(x),~\theta_{7}(y)p_{0}^{(7)}(x)),\\
 \theta_{8}(y)p_{0}^{(7)}(x), \theta_{9}(y)p_{0}^{(9)}(x), \theta_{10}(y)p_{0}^{(9)}(x),~\theta_{11}(y)p_{0}^{(11)}(x),~\theta_{11}(y)p_{1}^{(11)}(x)\}$ generates a 2D cyclic code over $\mathfrak{R}$.
\end{example}
\begin{example}
Consider the finite chain ring $\mathfrak{R}= \mathbb{F}_{17}+\gamma \mathbb{F}_{17};~ \gamma^{2}=0$. Let $\mathcal{C}$ be 2D cyclic code of length $mn$ over $\mathfrak{R}$, where $m=17~ \text{and} ~n=4$. Then $\mathcal{C}$ can be viewed as an ideal of $R=\mathfrak{R}[x,y]/\langle x^{17}-1,~y^{4}-1\rangle$. It can be easily seen that $4^{\gamma}(mod~289)$ is primitive $4^{th}$ root of unity in $\mathfrak{R}$. Therefore, by Lemma \ref{lem.1}, $\theta_{j}(y)= \frac{1}{4}(1+4^{\gamma(4-j)} y+4^{2\gamma(4-j)}y^{2} + 4^{3{\gamma(4-j)}}y^{3})$ are primitive central idempotents of $\mathfrak{R}[y]/\langle y^{4}-1\rangle;~ 0\leq j\leq 3$. Consider the following cyclic codes of length $17$ over $\mathfrak{R}$
\begin{align*}
    I_{0}&=\langle p_{0}^{(0)}(x)\rangle= \langle \gamma(x^{3}-3x^{2}+3x-1)\rangle\\ 
    I_{1}&=\langle p_{0}^{(1)}(x)\rangle=\langle x^{6}-6x^{5}+15x^{4}-20x^{3}+15x^{2}-6x+1\rangle\\
     I_{2}&=\langle p_{0}^{(2)}(x),~p_{1}^{(2)}(x)\rangle\\
     &=\langle \gamma(x^{5}-5x^{4}+10x^{3}-10x^{2}+5x-1),\\
     &~~~~~x^{7}-7x^{6}+21x^{5}-35x^{4}+35x^{3}-21x^{2}+7x-1+\gamma(x^{3}+2x^{2}-3x+2) \rangle\\
    I_{3}&=\langle p_{0}^{(3)}(x),~p_{1}^{(3)}(x)\rangle\\
    &=\langle \gamma(x^{4}-4x^{3}+6x^{2}-4x+1),\\
    &~~~~~x^{9}-9x^{8}+36x^{7}-84x^{6}+126x^{5}-126x^{4}+84x^{3}-36x^{2}+9x-1 \rangle
\end{align*}
 By Theorem \ref{th.5}, the set $\{ \theta_{0}(y)p_{0}^{(0)}(x),~\theta_{1}(y)p_{0}^{(1)}(x),~\theta_{2}(y) p_{0}^{(2)}(x),~\theta_{2}(y)p_{1}^{(2)}(x),\\~\theta_{3}(y)p_{0}^{(3)}(x),~\theta_{3}(y)p_{1}^{(3)}(x)\}$ generates a 2D cyclic code over $\mathfrak{R}$.
\end{example}

\section{An Algorithm to Find the Generators of an $n$D Cyclic Code}\label{sec4}

An algorithm to find the generators of an $n$D cyclic code viewed as an ideal of $R_{n}=\mathfrak{R}[x_{1},x_{2},\dots,~x_{n}]/\langle x_{1}^{m_{1}}-1,~x_{2}^{m_{2}}-1,\dots,~x_{n}^{m_{n}}-1\rangle$ over a finite chain ring $\mathfrak{R}$ with residue field $\mathbb{F}_{q}$ is given below:\\
    \textbf{Step 1.}
    1D cyclic codes of length $m_{1}$ over $\mathfrak{R}$ are ideals of $R_{1}=\mathfrak{R}[x_{1}]/\langle x_{1}^{m_{1}}-1\rangle $ and their generators can be determined by Theorem \ref{th.1}.\\
    \textbf{Step 2.} 2D cyclic codes of length $m_{1}m_{2}$ over $\mathfrak{R}$ are ideals of
    \begin{align*}
        R_{2}=\mathfrak{R}[x_{1},~x_{2}]/\langle x_{1}^{m_{1}}-1,~x_{2}^{m_{2}}-1\rangle &\cong (\mathfrak{R}[x_{1}]/\langle x_{1}^{m_{1}}-1\rangle)[x_{2}]/\langle x_{2}^{m_{2}}-1\rangle\\
        &\cong R_{1}[x_{2}]/\langle x_{2}^{m_{2}}-1\rangle.
    \end{align*}
   Using the generators of ideals of $R_{1}$ obtained in Step 1, generators of such codes can be determined by Theorem \ref{th.2}, if $m_{1},~m_{2}$ are arbitrary; and by Theorem \ref{th.5}, if $q\equiv 1(mod~ m_{2})$.\\
    \textbf{Step 3.} 3D cyclic codes of length $m_{1}m_{2}m_{3}$ over $\mathfrak{R}$ are ideals of
    \begin{align*}
        R_{3}=\mathfrak{R}[x_{1},~x_{2},~x_{3}]/\langle x_{1}^{m_{1}}-1,~x_{2}^{m_{2}}-1,~x_{3}^{m_{3}}-1\rangle \cong R_{2}[x_{3}]/\langle x_{3}^{m_{3}}-1\rangle.
    \end{align*}
    Generators of ideals of $R_{3}$ can be obtained by using the generators of ideals of $R_{2}$ obtained in Step 2 and thereafter using Theorem \ref{th.5} or Theorem \ref{th.2},  depending upon whether $q\equiv 1(mod~ m_{3})$ or not.\\
    \textbf{Step 4.} Continuing in same fashion, suppose we have obtained the generators of ideals of  $R_{n-1}$. Using the generators of ideals of $R_{n-1}$ and the fact that $R_{n}\cong R_{n-1}[x_{n}]/\langle x_{n}^{m_{n}}-1\rangle$, Theorem \ref{th.2} or Theorem \ref{th.5} can be applied to find the generators of ideals of $R_{n}$ for any natural number $n$.   

\begin{remark}
Note that while performing these iterations, we may interchange the role of $x_{i}'s$ to assume without loss of generality that $m_{1},m_{2},\dots,~m_{t}$ do not divide $q-1$ and $m_{t+1},m_{t+2},\dots,~m_{n}$ divide $q-1$. Then, we shall get the generators of an $n$D cyclic code over a finite chain ring in the form
\begin{align*}
    f(x_{1},x_{2},\dots,~x_{t})f_{t+1}(x_{t+1})\dots f_{n}(x_{n}).
\end{align*}

\end{remark}
 
\section{Conclusion}\label{sec13}

In this paper, the generator polynomials of a 2D cyclic code $\mathcal{C}$ of arbitrary length over a finite chain ring $\mathfrak{R}$ have been determined by two different methods. In Method 1, the generators of a 2D cyclic code of arbitrary length over $\mathfrak{R}$ have been determined by using generators of cyclic code of length $m$ over $\mathfrak{R}$. Method 2 is applicable to find the generators of a 2D cyclic code of length $mn$ such that either $m\vert q-1$ or $n\vert q-1$. This method gives the generator polynomials of $\mathcal{C}$ in separable form. Finally, for a natural number $n$, an algorithm based on both the methods for finding generators of an $n$D cyclic code has been developed.\\

\noindent \textbf{Acknowledgments}
The first author would like to thank University Grants Commission (UGC) New Delhi, India for financial support.

\end{document}